\documentclass[a4paper,fleqn]{cas-sc}

\usepackage[authoryear]{natbib}
\usepackage{graphicx} 
\usepackage{float}
\usepackage{algorithm}  
\usepackage{algpseudocode}
\usepackage{color}
\usepackage{setspace}

\newcommand{\pysimfrac}{{\sc pySimFrac} }
\newcommand{\vy}{{\mathbf y}}
\newcommand{\vx}{{\mathbf x}}

\usepackage{listings}
\definecolor{codegreen}{rgb}{0,0.6,0}
\definecolor{codegray}{rgb}{0.5,0.5,0.5}
\definecolor{codepurple}{rgb}{0.58,0,0.82}
\definecolor{backcolour}{rgb}{0.95,0.95,0.92}

\lstdefinestyle{mystyle}{
    backgroundcolor=\color{backcolour},   
    commentstyle=\color{codegreen},
    keywordstyle=\color{magenta},
    numberstyle=\tiny\color{codegray},
    stringstyle=\color{codepurple},
    basicstyle=\ttfamily\footnotesize,
    breakatwhitespace=false,         
    breaklines=true,                 
    captionpos=b,                    
    keepspaces=true,                 
    numbers=left,                    
    numbersep=5pt,                  
    showspaces=false,                
    showstringspaces=false,
    showtabs=false,                  
    tabsize=2
}

\newif\ifnotesw \noteswtrue

\def\tsc#1{\csdef{#1}{\textsc{\lowercase{#1}}\xspace}}
\tsc{WGM}
\tsc{QE}
\tsc{EP}
\tsc{PMS}
\tsc{BEC}
\tsc{DE}

\usepackage{lineno}

\begin{document}
\let\WriteBookmarks\relax
\def\floatpagepagefraction{1}
\def\textpagefraction{.001}
\shorttitle{\pysimfrac}
\shortauthors{Guiltinan}

\title [mode = title]{{\sc pySimFrac}: A Python Library for Synthetic Fracture Generation, Analysis, and Simulation}

\author[1]{Eric Guiltinan}[type=editor,
                        auid=000,bioid=1,orcid=0000-0002-0763-0625]
\credit{writing, code development, MF-LBM integration}

\author[1]{Javier E. Santos} 
\credit{writing, code development, MP-LBM integration}

\author[1]{Prakash Purswani}
\credit{writing, documentation, MF-LBM integration}

\author[1]{Jeffrey D. Hyman}
\credit{writing, code development, documentation, dfnWorks integration, and examples}

\address[1]{Earth and Environmental Science Division, Los Alamos National Laboratory, Los Alamos, NM, 87545}

\begin{abstract}
In this paper, we introduce \pysimfrac, a open-source python library for generating 3-D synthetic fracture realizations, integrating with fluid simulators, and performing analysis.  \pysimfrac allows the user to specify one of three fracture generation techniques (Box, Gaussian, or Spectral) and perform statistical analysis including the autocorrelation, moments, and probability density functions of the fracture surfaces and aperture.  This analysis and accessibility of a python library allows the user to create realistic fracture realizations and vary properties of interest.  In addition, \pysimfrac includes integration examples to two different pore-scale simulators and the discrete fracture network simulator, dfnWorks. The capabilities developed in this work provides opportunity for quick and smooth adoption and implementation by the wider scientific community for accurate characterization of fluid transport in geologic media. We present \pysimfrac along with integration examples and discuss the ability to extend \pysimfrac from a single complex fracture to complex fracture networks. 
\end{abstract}

\maketitle 

\printcredits

\doublespacing

\section{Introduction}
\label{intro}
The study of complex fracture geometries has important applications in Earth and material sciences.  Fractures are ubiquitous in geologic formations where they represent preferential flow pathways in otherwise low permeable materials~\citep{viswanathan2022from}.  These conduits often control the response of fluid migration in the subsurface which has important implications for unconventional oil and gas exploration, geothermal energy, environmental remediation, carbon dioxide sequestration, hydrogen and natural gas storage, and nuclear waste isolation \citep{renshaw1995relationship, wang2014non, wang2015modification, vogler2018experiments}. Experimental and numerical work in natural fractures is often challenging due to the difficulty of obtaining samples representative of subsurface conditions and the inability to model the wide range of relevant length scales, which span multiple orders of magnitude~\citep{bonnet2001scaling}.   To bridge the gap between natural fractures available for experimentation and the large variety of expected fractures researchers often turn to synthetic fracture generation techniques \citep{Brown_1995_Simple}.

Several synthetic fracture generation techniques have been developed \citep{Ogilvie_2006_AUPG_SynFrac,Brown_1995_Simple,Glover_1998_Fluid,Glover_1998_Synthetic,Glover_1999_Characterizing}.  \cite{Brown_1995_Simple} presented a Fourier space-based mathematical model for the generation of synthetic fracture surfaces which relied upon only three parameters; the fractal dimensions, the roughness, and a mismatch length scale.  This model assumes that at lengths less then the mismatch length the two fracture surfaces are completely uncorrelated and at lengths greater than the mismatch length they are perfectly matched.  \cite{Glover_1998_Fluid} presented a more realistic model which included a transition length where the fracture surfaces transitions smoothly from completely uncorrelated to a specified maximum correlation.  \cite{Ogilvie_2006_AUPG_SynFrac} presented an update to the \cite{Glover_1998_Fluid} method, which corrected an error in the the mixing of correlated random variables and also included the ability to specify a minimum correlation at short length scales.  The techniques discussed here have been implemented in a graphical user interface based program called "SynFrac" \citep{Ogilvie_2006_AUPG_SynFrac}.  However, SynFrac has some limitations.  In particular, it can only generate fractures with square dimensions (e.g. 128x128, 256x256), only outputs csv or txt files, has limited analysis tools, and cannot be called within automated scripts.  This makes the development of large datasets of fracture properties (e.g., \cite{Guiltinan_2021, en15238871}) time consuming and prone to errors.
Moreover, many research teams have developed one-off scripts to generate synthetic fracture surfaces but there is not a comprehensive open source scripted toolkit available at this time. 

To overcome the limitations in currently available fracture generation methods, we have developed \pysimfrac. \pysimfrac is a Python module for constructing 3D single fracture geometries. The software is designed to help researchers investigate flow through fractures through direct numerical simulations of single/multi-phase flow through fractures. One advantage of the python implementation is that it allows for greater flexibility and customization compared to a GUI-based approach. With a python-based interface, researchers can readily expand the development and test new fracture generation algorithms or modify existing methods to better match experimental data. \pysimfrac offers spectral-based and convolution-based fracture generation methods. Both methods can be customized to produce synthetic fractures akin to different rock types.  
\pysimfrac also includes utilities for characterizing surface and aperture properties such as the correlation length, moments, and probability density function of the fracture surfaces and aperture field. 

\pysimfrac also provides seamless integration with open-source flow simulation libraries (MF-LBM\citep{CHEN201814}, MP-LBM~\citep{mplbm}, and {\sc dfnWorks}~\citep{hyman2015dfnworks}) elevating its utility for researchers and practitioners alike. This ease of integration streamlines the process of conducting direct numerical simulations of single/multi-phase flow through fractures, fostering a comprehensive understanding of fluid dynamics within these complex structures. By providing built-in compatibility with popular open-source simulators, \pysimfrac eliminates the need for time-consuming and error-prone manual configuration, allowing users to focus on their research objectives. The library's robust and extensible design caters to a wide array of applications, accommodating users with varying requirements and expertise. Ultimately, \pysimfrac's integration with flow simulation libraries further enhances its value as a tool for investigating fracture flow behavior, contributing significantly to advancements in subsurface hydrology, reservoir engineering, and environmental studies.

\section{Software Design}

\pysimfrac has three primary components: (1) fracture surface generation (2) analysis toolkit and (3) and interface with various flow and transport solvers via file formatting. \pysimfrac is meant to be an interactive object oriented module a primary class for a single fracture. The generation method along with functions to obtain statistical information, visualization/plotting, and input/output are all attached to the object. 

\begin{figure}
    \centering
    \includegraphics[width=0.70\textwidth]{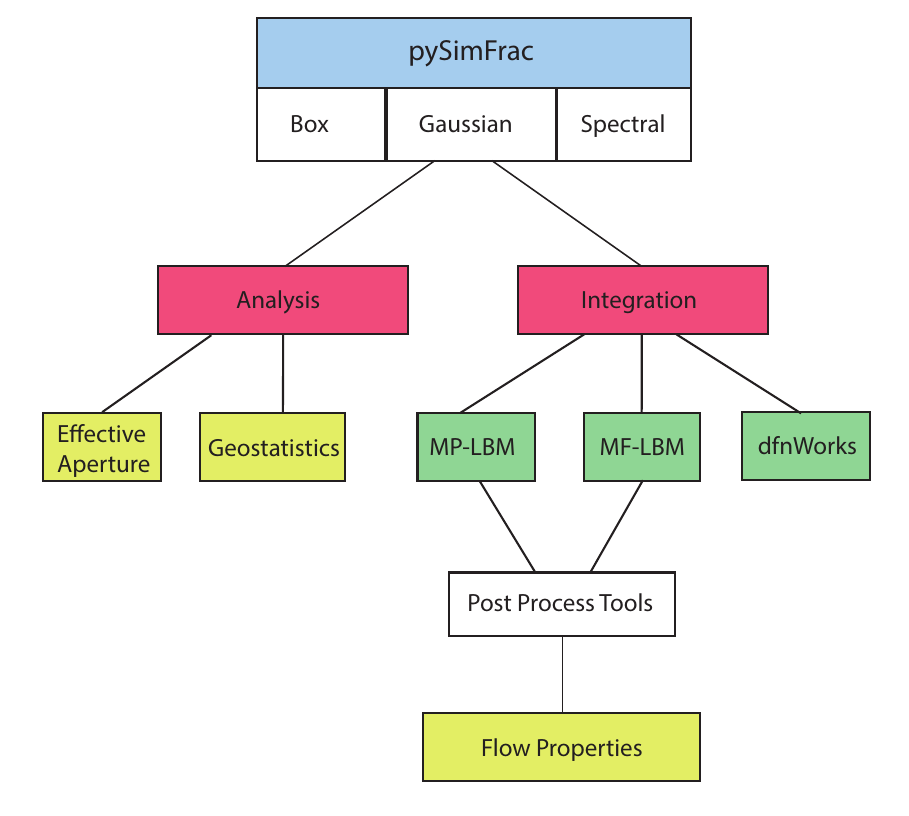}
    \caption{\pysimfrac contains three fracture generation methods (Box, Guassian, and Spectral) as well as analysis and integration with flow simulators.}
    \label{fig:pysimfrac}
\end{figure}

\subsection{Fracture Surface Generation Methods}

\pysimfrac has multiple generation methods to produce rough fracture surfaces. 
The methods can be broken into two primary categories. The first set of techniques are spectral methods that produce self-affine / fractal surfaces. The second set are convolution-based. In addition to generating synthetic fractures, \pysimfrac can be used to read in profilometry of fracture surfaces obtained in the laboratory.

Each \pysimfrac object instance is defined by its length in x ($l_x$) and y ($l_y$) and a discretization length $h$. 
Therefore, a uniform sized grid is produced with $nx = \lceil lx/h \rceil$ and $ny = \lceil lx/h \rceil$ for the top and bottom fracture surfaces and the projected aperture field. The latter is the difference in the heights between the two surfaces at each discrete location in the grid.  \pysimfrac allows for specifying a mean aperture field which is controlled during the voxelization process.  Along with these domain parameters, the user must specify a generation method, which are detailed below. 

\subsubsection{Spectral Method}

Rough fracture surfaces have been represented by fractal / self-affine models in numerous studies~\citep{da2019impact,kang2016emergence,stigsson2019novel,wang2016influence}. At a first order approximation, The Fourier decomposition of a rough surface indicates that many non-uniform surfaces exhibit a power-law decay in the power spectral density function with a function form of
\begin{linenomath*}
\begin{equation}\label{eq:spectral_decay}
G(k) = C k^{-\alpha}
\end{equation}
\end{linenomath*}
where $k = 2 \pi/\lambda$ is the wave number / Fourier mode, $\lambda$ is the wavelength, $C$ is a proportionality constant, and $\alpha$ is the decay exponent.  Based on these observations, a number of spectral / Fourier based rough surface generation methods have been proposed, the most common being~\citet{Brown_1995_Simple}, \citet{Glover_1998_Synthetic}, and~\citet{Ogilvie_2006_AUPG_SynFrac}.  A spectral method coded in matlab and based upon these techniques is available (https://github.com/rvillamor/digital\_generation
\_of\_fractures/blob/main/RSG\_brown1995.m) and these techniques can also be found in the synFrac program \newline 
(https://homepages.see.leeds.ac.uk/~earpwjg/PG\_EN/Images/Software/Manual\%20for\%20web/Create.htm).

While there are differences and chronological improvements between these method, the core portion of the algorithms are fairly consistent. The methods all modify the amplitude and phases of the Fourier components of the surfaces. The amplitudes are scaled according to~\eqref{eq:spectral_decay} and the phases are controlled using streams of random numbers. Special care is taken to define the random numbers which define phase, cf.\citet{Ogilvie_2006_AUPG_SynFrac} for a detailed discussion. The desired fractal dimension and autocorrelation of the surface is often defined in terms of the Hurst exponent which is in a particular sense related to $\alpha$ in~\eqref{eq:spectral_decay}. These features along with anisotropy are included into the method via the amplitudes of the decomposition. The spectral method implemented in \pysimfrac has the following parameters: (1) Hurst exponent with range; range (0,1), (2) Roughness / standard deviation of heights; range $\ge 0$, (3) Anisotropy ratio; range (0,1), (4) $\lambda_0$ roll-off length scale as a fraction of fracture size; range [0,1], (5) Mismatch length scale (wavelength) as a fraction of fracture size [0,1] (6) Power spectral density model roll-off function (linear / bilinear / smooth).

\begin{figure}
    \centering
    \includegraphics[width = 0.9\textwidth]{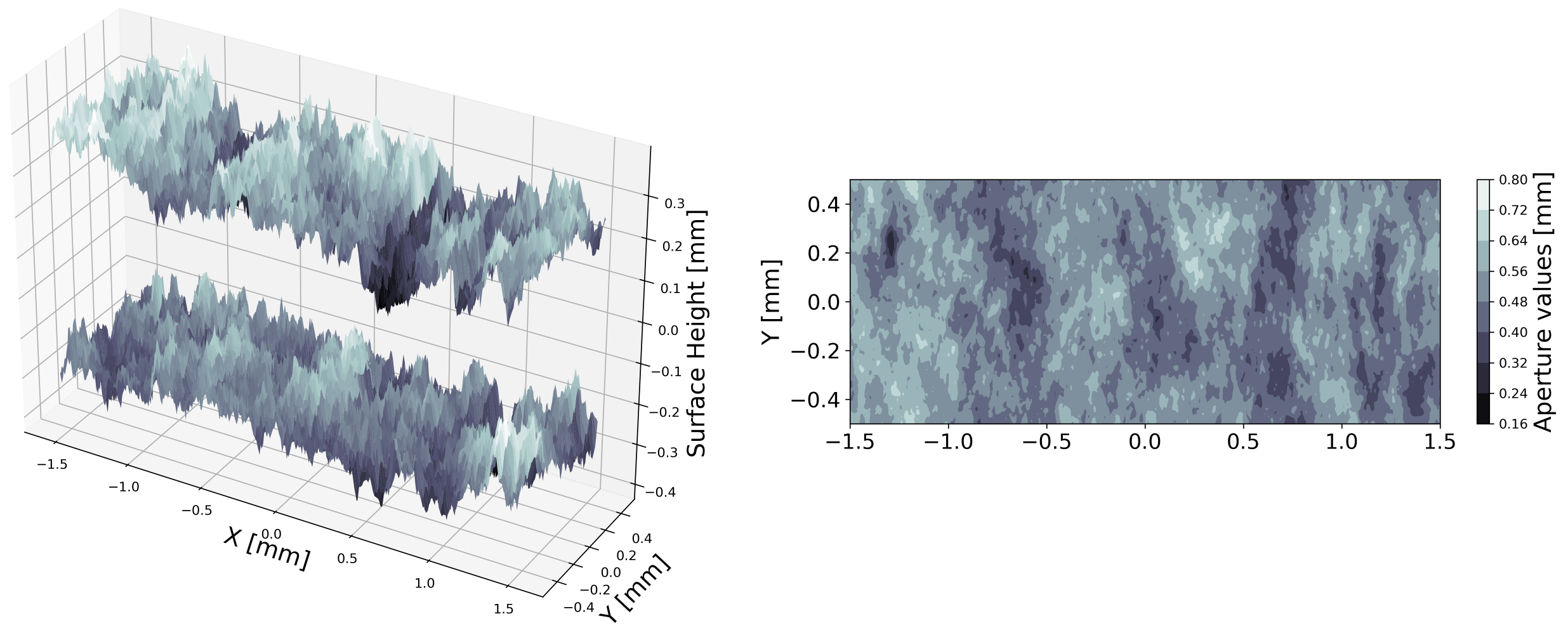}
    \caption{Fracture surface generated using the convolution method with an anisotropic Gaussian kernel   \label{fig:spectral}}
\end{figure}

An example of a surface generated using the spectral method is provided in Fig.~\ref{fig:spectral} and the code used to generate it is shown in listing~\ref{code:spectral}.
The top and bottom surfaces are shown on the left sub-figure and the projected aperture field is in the right sub-figure. 
The sample was generated with a mean aperture of 0.5 mm, a roughness of 0.5, and an anisotropy of 0.5, mismatch of 0.1, and the smooth power spectral density model.
\begin{lstlisting}[language=Python, label=code:spectral, caption=Generation of fracture surface object using the \pysimfrac spectral method.]
Spectral = SimFrac(
                    h=0.01,
                    lx=3,
                    ly=1,
                    method="spectral",
                    units='mm'
                )

Spectral.params["H"]["value"] = 0.5
Spectral.params["mean-aperture"]["value"] = 0.5,
Spectral.params["roughness"]["value"] = 0.05
Spectral.params["aniso"]["value"] = 0.5
Spectral.params["mismatch"]["value"] = 0.1
Spectral.params["model"]["value"] = "smooth"

\end{lstlisting}

\subsubsection{Convolution Methods}
The convolution methods are based on creating a stationary random topography by convolving an uncorrelated random field ($u(\vx) \sim U[0,1]$) with a specified kernel ($k(\vx)$)
\begin{linenomath*}
\begin{equation}\label{eq:convolution}
    T(\vx) = \int d \vy~k(\vx - \vy) \ast u(\vx)~\,.
\end{equation}
\end{linenomath*}
The structure  of the $T(\vx)$ (moments, correlation, and anisotropy)  are determined by the central limit theorem and the inheriated properties of the kernel.
\pysimfrac has several built in kernels, with the primary being a multi-variant 2-dimensional Gaussian function of the form
\begin{equation}
    k(\vx) = \frac{1}{2 \pi \sqrt{Det(\Lambda)}}\exp\left [ -\vx^\prime \Lambda \vx/2 \right]\,,
\end{equation}
where $\Lambda$ is symmetric matrix of length scales whose elements $\lambda_i$ determine the spread of $k(\vx)$ in various directions.
Equation~\eqref{eq:convolution} produces a single surface topography with mean 0 and variance determined by the support of $k(\vx)$, a direct result of the central limit theorem~\cite{Hyman_2014}.
Thus, to produce a fracture with desired mean aperture and variance, a copy of $T(\vx)$ is created as the bottom surface, then both surfaces translated and rescaled to obtained to the desired values.
Isotropic topographies can be created by defining $\Lambda$ as a diagonal matrix and assigning the same length scale $\lambda$ to every direction.
Anistropic ones, but having the valus unequal, i.e., larger values in $\lambda_x$ than $\lambda_y$ will create longer correlations in the $x$-direction that $y$-direction.
\citet{Hyman_2014} introduced this method for the generation of explicit three-dimensional pore structures, which has found use in various applications and studies~\citep{guedon2017influence,hyman2012heterogeneities,hyman2013hyperbolic,hyman2013pedotransfer,hyman2015statistical,siena2015direct,siena2014relationship}.
The co-variance function as well as other properties of $T(\vx)$ generated with the Gaussian kernel are given explicitly in~\citet{Hyman_2014}.
Its worth noting that the surface generated using the Gaussian kernel are infinitely smooth in the mathematical sense because the smoothness (infinitely differentiable) is transferred to the surface via the convolution. 
In addition to the Gaussian kernel, there is a uniform or box function kernel available in \pysimfrac, and the inclusion of additional kernels is straightforward and an area of active development.

\begin{figure}
    \centering
    \includegraphics[width = 0.9\textwidth]{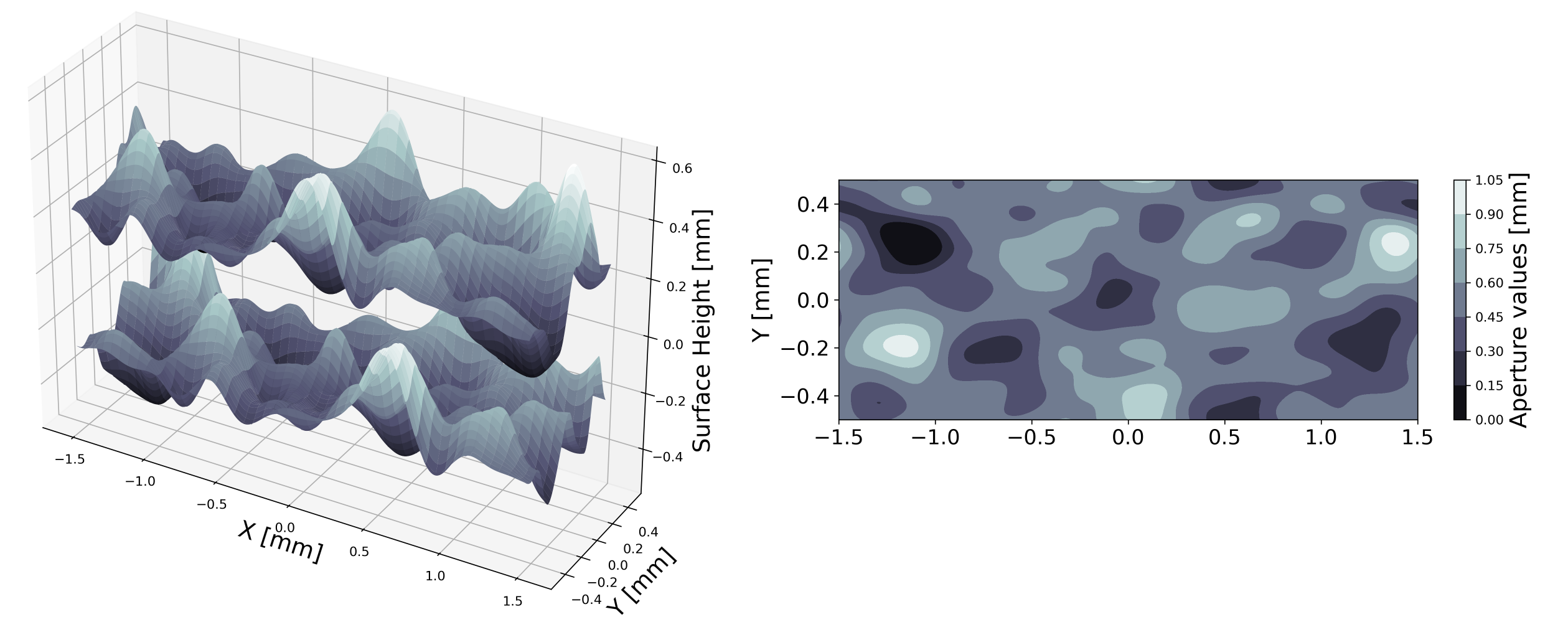}
    \caption{Fracture surface generated using the convolution method and the Gaussian kernel   \label{fig:gaussian}}
\end{figure}

An example of a surface generated using the convolution method and the Gaussian kernel is shown in Fig.~\ref{fig:gaussian} and the code shown in listing~\ref{code:gaussian}.
The top and bottom surfaces are shown on the left sub-figure and the projected aperture field is in the right sub-figure. 
The sample was generated with a mean aperture of 0.5 mm, a log variance of 0.01, and an anisotropic kernel ($\lambda_{1,1} = 0.15$, $\lambda_{2,2} = 0.25$), and a shear of 0.5 mm, which translates the top surface along the x-axis 0.5 mm to mimic shear (additional details provided in the next sections).

\begin{lstlisting}[language=Python, label=code:gaussian, caption=Generation of fracture surface object using the \pysimfrac convolution method with a Gaussian kernel.]
Gaussian = simfrac(
    method = "gaussian",
    h = 0.01, 
    lx = 3, 
    ly = 1)
    
Gaussian.params["mean-aperture"]["value"] = 0.5
Gaussian.params["aperture-log-variance"]["value"] = 0.01
Gaussian.params["lambda_x"]["value"] = 0.15
Gaussian.params["lambda_y"]["value"] = 0.25
Gaussian.shear = 0.5
Gaussian.create_fracture()
\end{lstlisting}

\subsection{Additional Generation Functions}
\begin{figure}
    \centering
    \includegraphics[width = 0.9\textwidth]{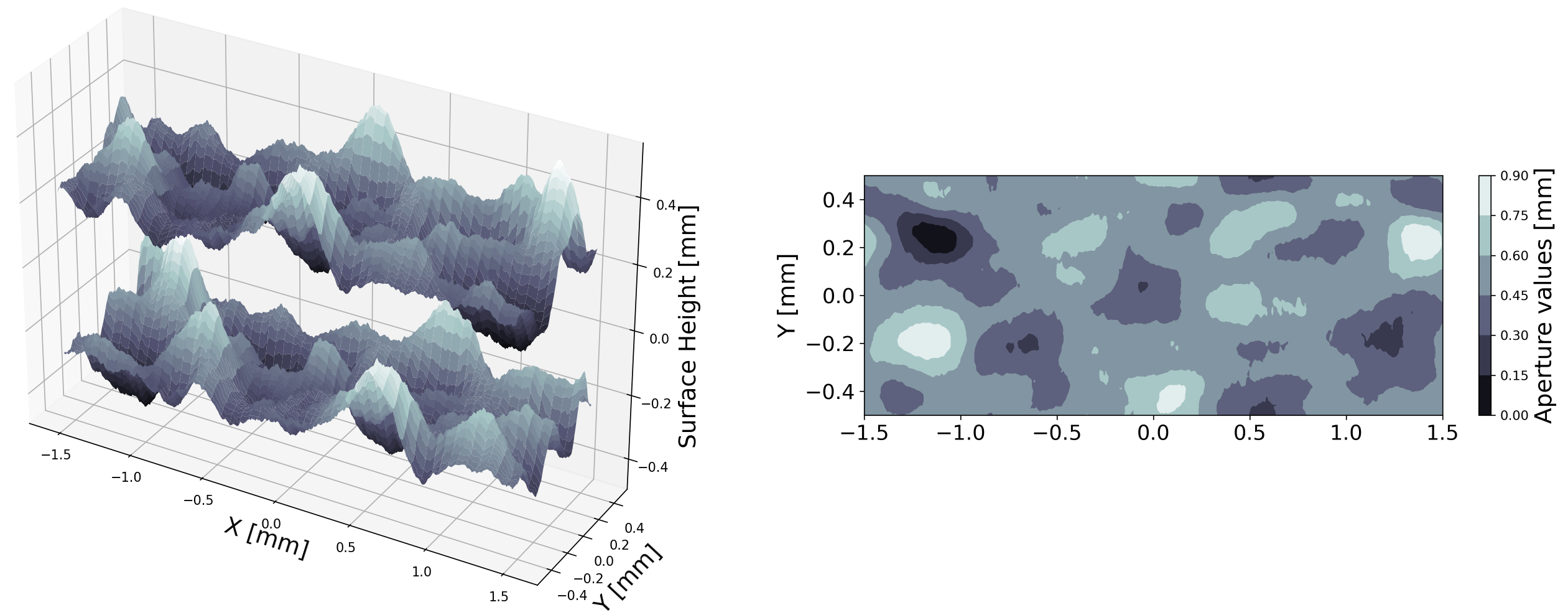}
    \caption{Combined fracture surface of Fig.~\ref{fig:spectral} and Fig.~\ref{fig:gaussian}.  \label{fig:combined}}
\end{figure}

In addition to the base generation methods detailed above, there are a number of functions in \pysimfrac to further manipulate the surfaces. Foremost, one can rescale the mean and variance of the surfaces, jointly or individually, and the mean projected aperture field using any desired value. Next, one can apply horizontal shear to the fracture by shifting the top fracture surface along the x-axis for the desired distance. A key property of the \pysimfrac fractures is that they are periodic in both x and y and the shear effectively translates the surface around a torus. Thus, the shear translation does not introduce discontinuities in the surfaces nor shorten the domain size, which could be the case if the surface was not periodic. Maintaining periodicity in x and y is often an important requirement of numerical simulation, particularly when simulating steady state fluid distributions for relative permeability calculations. Finally, \pysimfrac surfaces can be combined using weighted linear superposition to create new surfaces. 
An example of this is shown in Fig.~\ref{fig:combined} and listing~\ref{code:combined}. 
Here, we combined the surfaces shown in Fig.~\ref{fig:spectral} and Fig.~\ref{fig:gaussian} with 1/4 and 3/4 weights, respectively. The resulting fracture surface inherits the long correlations from the Gaussian kernel convolution surface as well as the local roughness of the spectral method. Any number of fracture objects can be combined.

\begin{lstlisting}[language=Python, label=code:combined, caption=Combination of multiple fracture objects.]
## Create new fracture object that is the weighted linear superposition of two existing surfaces (See previous listings)
Combined = Spectral.combine_fractures([Gaussian], weights = [0.25, 0.75])
## Plots fracture surfaces     
Combined.plot_surfaces() 
## Plots fracture aperture     
Combined.plot_aperture_field()
\end{lstlisting}

Finally, \pysimfrac also allows users to import surfaces obtained from real fracture scans using profilometry so long as they are mapped onto a regular grid with equal spacing in both directions.
Preprocessing of the raw profilometry is not supported as part of \pysimfrac module. 

\subsection{Analysis tools}

In addition to the generation of fracture surfaces, \pysimfrac provides a suite of geostatistical analysis tools. 
Functions are included to compute the first four moments of the surface height and aperture distributions.

\begin{lstlisting}[language=Python, label=code:analysis, caption=Computation of geostatical analysis]
## Compute first four moments of the surface height and aperture distributions
Spectral.compute_moments()
## Plot Height and aperture PDFs
Spectral.plot_surface_pdf()

\end{lstlisting}

\begin{figure}
    \centering
    \includegraphics[width = 0.9\textwidth]{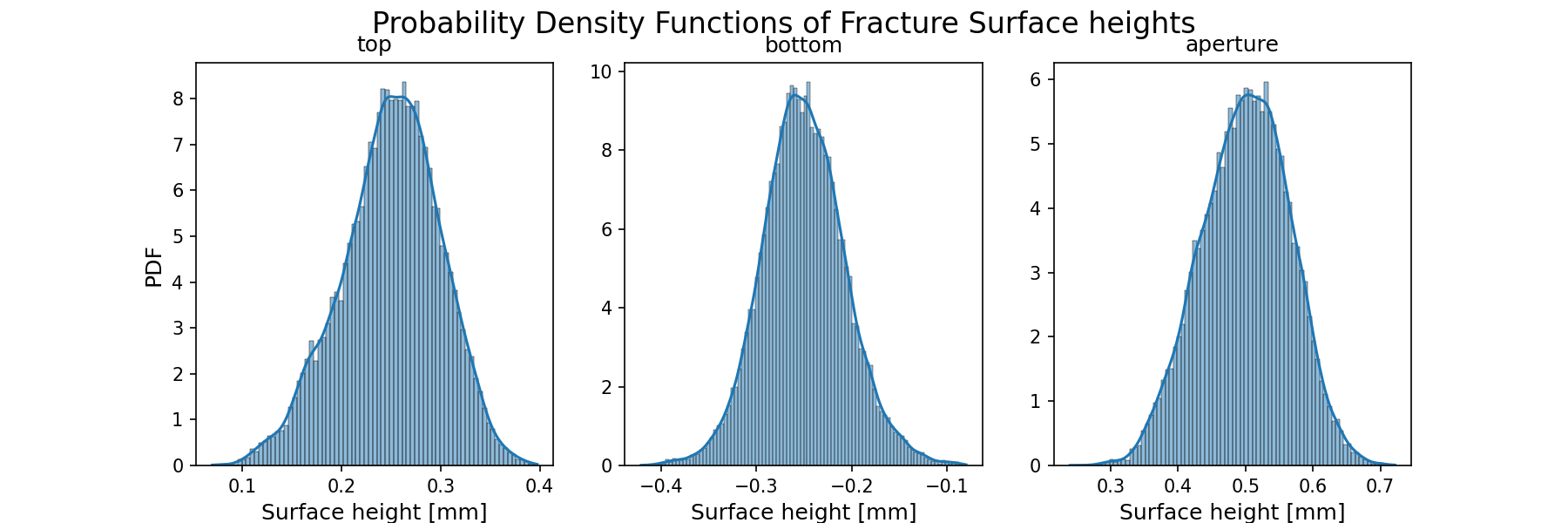}
    \caption{Probability density functions of the surface shown in Fig.~\ref{fig:spectral}  \label{fig:spectral_pdf}}
\end{figure}

\begin{figure}
    \centering
    \includegraphics[width = 0.9\textwidth]{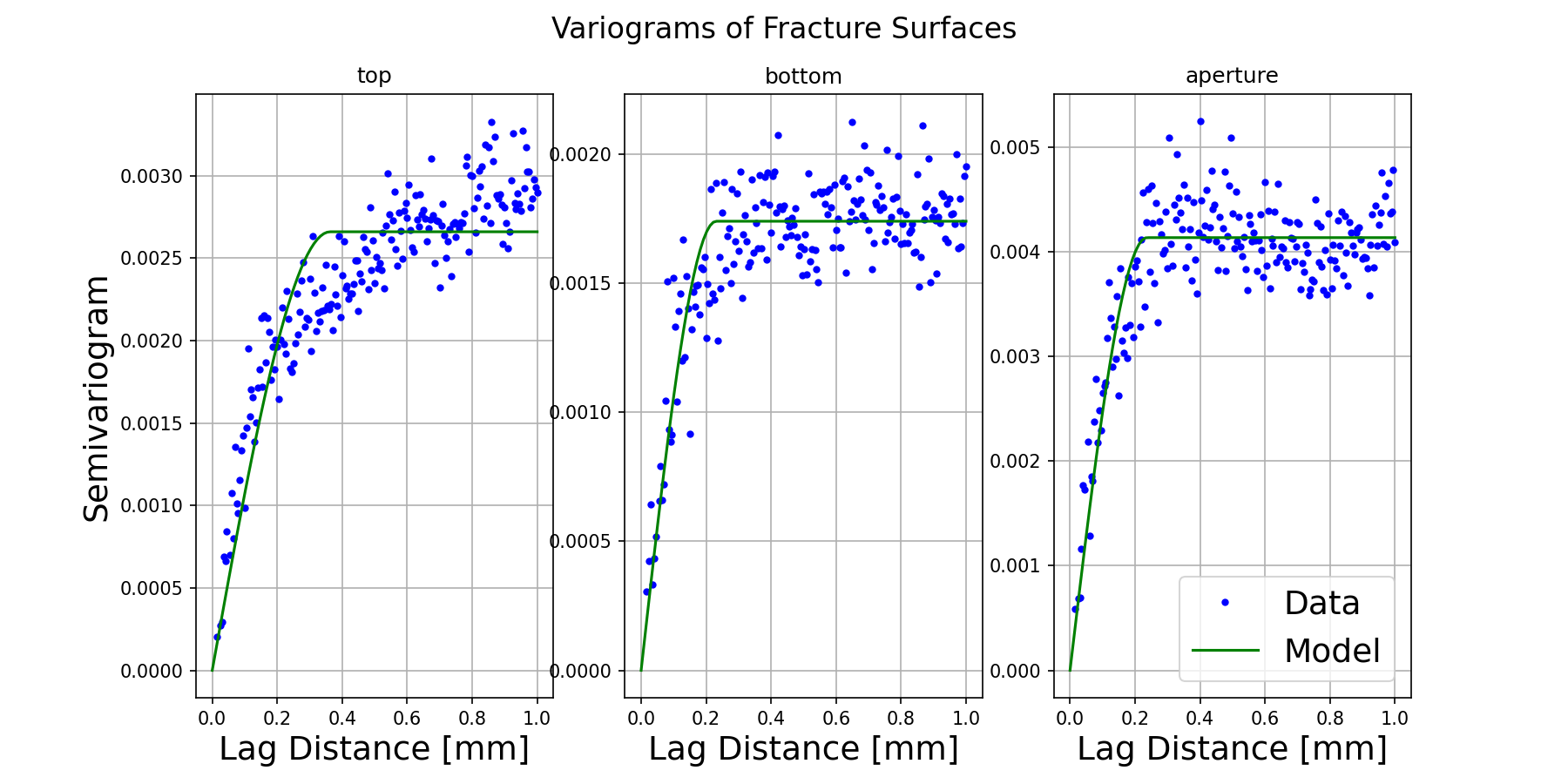}
    \caption{Semi-variogram with spherical model for the surface shown in Fig.~\ref{fig:spectral}  \label{fig:spectral_variogram}}
\end{figure}

In addition to the moments of the height and aperture distributions, functions are included to compute and plot the auto-correlation function of the surface in the principal Cartesian directions (x and y) and the radial semi-variogram. 
The semi-variogram is computed by calling the python module {\sc SciKit-GStat}~\citep{malicke2022scikit}. 
Figure~\ref{fig:spectral_variogram} and Listing~\ref{code:variogram} provide an example for the surface shown in Fig.~\ref{fig:spectral}.








\begin{lstlisting}[language=Python, label=code:variogram, caption=Computation of geostatical analysis]
Spectral.compute_variogram(max_lag = 100, num_lags = 200)
Spectral.plot_variogram()
\end{lstlisting}

\subsection{Effective Properties}
Estimations of the effective properties from the structure of the surfaces are also provided by \pysimfrac. These estimations are categorized into two types. The first are analytical and empirically derived estimations of the effective hydraulic aperture and the second are numerical approximations.  The first estimations includes standard approximations such as various means, e.g., arithmetic, harmonic, and geometric, as well as several models proposed in the literature, cf.~\citet{he2021corrected} for a comprehensive list of models proposed in the literature.  Most of the models proposed in the literature use moments of the aperture distribution, which can be directly computed using the analysis toolkit. In principle, any effective hydraulic model with geo-statistical parameters can be added to \pysimfrac. The second type of approximations are obtained by numerical inversion of the Darcy equation with a spatially variable permeability field $k(\vx)$ inferred using a local cubic law from the aperture field $b(\vx)$, that $k(\vx) = b^2(\vx)/12$. Note, that other functional relationships between $k(\vx)$ and $b(\vx)$ can be readily applied as well.  We obtain pressure and volumetric flow rates by solving the standard flow equations with variable coefficients discretized using a second-order finite scheme.  Flow is driven by Dirchelet pressure boundary conditions in one primary direction to obtain estimates of effective permeability in that direction, which is then converted to an effective hydraulic aperture. 

%







%

\section{Integration with Flow and Transport Simulators}
In addition to the generation methods and analysis toolkits, we have developed seamless handoffs with several open-source flow and transport simulators ranging from multi-phase lattice Boltzman methods to three-dimensional discrete fracture network simulators. 

\subsection{Integration with MultiPhase LBM (MP-LBM)}

\begin{figure}
    \centering
    \includegraphics[width = 0.8\textwidth]{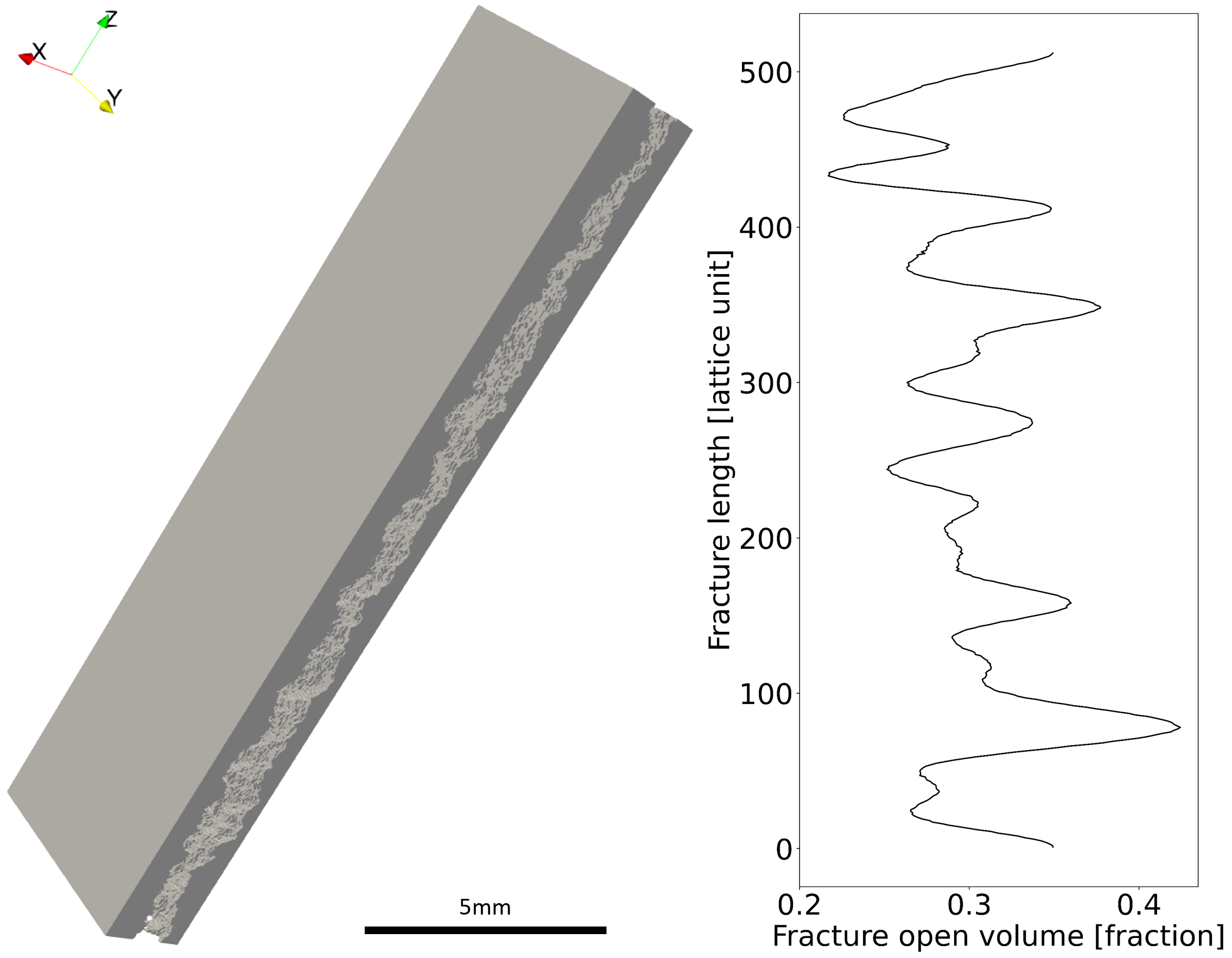}
    \caption{Characterizing the pore space of a typical fracture generated using \pysimfrac. The top-view of first slice for the fracture is shown on the left, while the porosity along the length of the fracture is shown on the right. The scale is adapted from the fracture mean aperture (0.584 mm) in \cite{karpyn2007visualization} }
    \label{Fig_4_1_1}
\end{figure}

To demonstrate the integration with flow simulators a fracture was created using \pysimfrac (Figure \ref{Fig_4_1_1}).  The fracture is 128 x 512 voxels and was created using the spectral method with the following parameters: \emph{model} 'smooth', \emph{roughness} 4.0, \emph{H} 0.7, \emph{aniso} 0.0, \emph{mismatch} 0.25, and \emph{mean aperture} 15. MP-LBM (\cite{mplbm}; https://
github.com/je-santos/MPLBM-UT) is a specialized lattice-Boltzmann library that significantly simplifies the process of running pore-scale lattice Boltzmann simulations through complex  porous media. MP-LBM uses the high-performance, highly parallel library Palabos (https://gitlab.com/unigespc/palabos) as the solver backend, making it easily deployable in a variety of systems, from laptops to supercomputer clusters. The \pysimfrac module aims to facilitate seamless integration between complex fracture geometry generation and single-phase flow simulation to enable the study of how realist fracture heterogeneities affect the permeability of a fracture domain. After creating a \pysimfrac object, and installing MP-LBM the simulation can be run by simply calling the write\_MPLBM function and supplying the number of buffer layers for voxelization, the number of cpus to be utilized, and number of computational hours requested (listing 6).

\begin{lstlisting}[language=Python, label=code:mplbm, caption=An example script demonstrating the integration with the MP-LBM code.] 
from wrappers import write_MPLBM, postprocess

# send simulation
lbm = write_MPLBM(simfrac_object = Spectral,
                  buffer_layers = 2,
                  cpus = 4,
                  num_hrs = 1)
                  
# plot and obtain permeability
postprocess(lbm)

\end{lstlisting}

\begin{figure}
    \centering
    \includegraphics[width=1.0\textwidth]{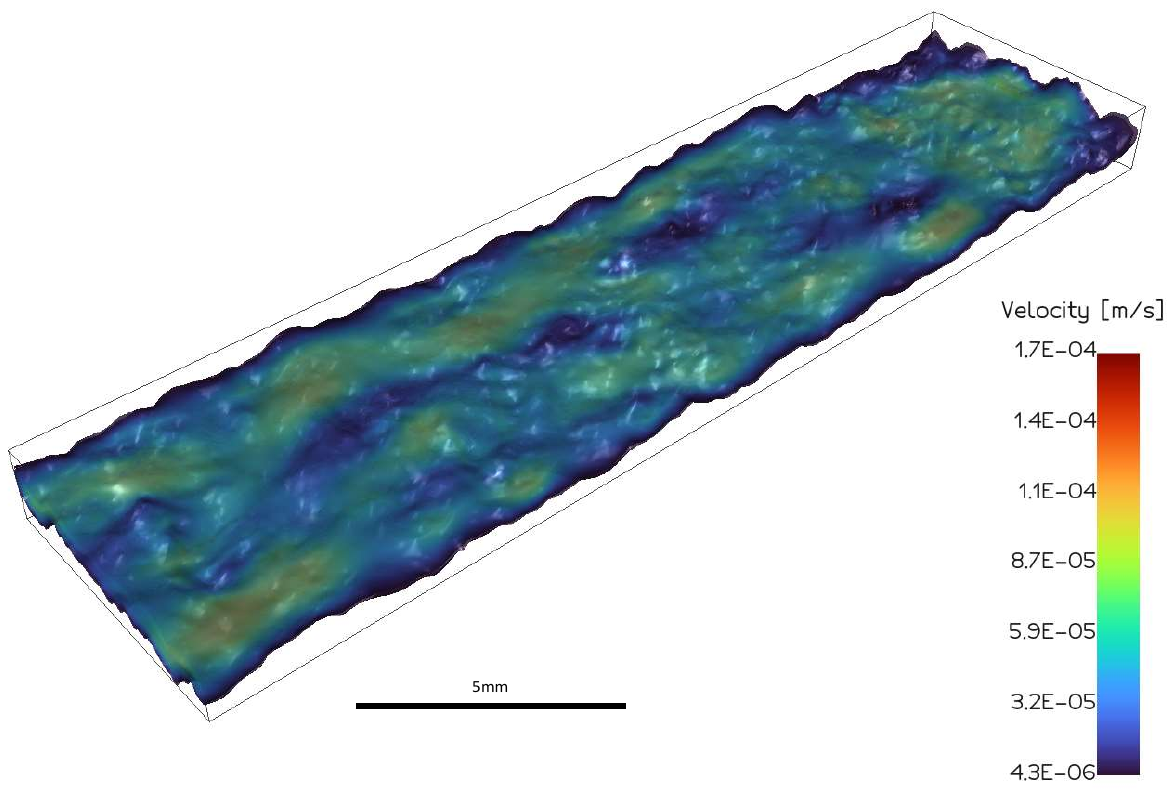}
    \caption{3D velocity field magnitude from a single-phase LBM Simulation performed using the MP-LBM extension. The depicted computational domain of the fracture measures 512x128 in the XY-plane, with a mean aperture of 12 voxels, corresponding to dimensions of 1.53e-6, 0.384e-6, and 0.045e-6 meters, respectively.}
    \label{fig:MPLBM}
\end{figure}

An example of the resulting flow field using the integration with MP-LBM is shown in Figure \ref{fig:MPLBM}.
The completion of a full simulation utilizing this feature required roughly 15 seconds of computing time on a standard laptop. We anticipate that the incorporation of this capability will streamline the research process, making various aspects of investigation more straightforward and productive for both researchers and practitioners. Understanding the relationship between geometry and permeability may offer innovative perspectives, potentially leading to the refinement of correlations that can be applied at the field scale. Furthermore, it could shed light on the effects of phenomena such as compaction, cementation, and dissolution on this critical parameter.

\subsection{Integration with MF-LBM}

MF-LBM \citep{CHEN201814,CHEN2019} is an open source (https://github.com/lanl/MF-LBM) high-performance lattice Boltzmann code developed at Los Alamos National Laboratory. It  combines the continuum-surface-force based color-gradient multiphase model \citep{Gunstensen_colorgradient_1991, Liu_colorgradient_2012} with the geometrical wetting model \citep{Leclaire_2016,Leclaire_2017, AKAI201856}.  The code is extensively parallelized and optimized for CPUs and GPUs and is ideal for running large (billion of nodes) multiphase simulations. 
We have integrated MF-LBM within \pysimfrac allowing users to not only specify fracture properties as part of the core \pysimfrac but also to specify multiphase flow parameters such as viscosity ratios, capillary number, contact angle, and interfacial tension. This allows for users to seamlessly conduct simulations on a variety of fracture properties with varied simulation parameters.

\begin{figure}
    \centering
    \includegraphics[width = 0.8\textwidth]{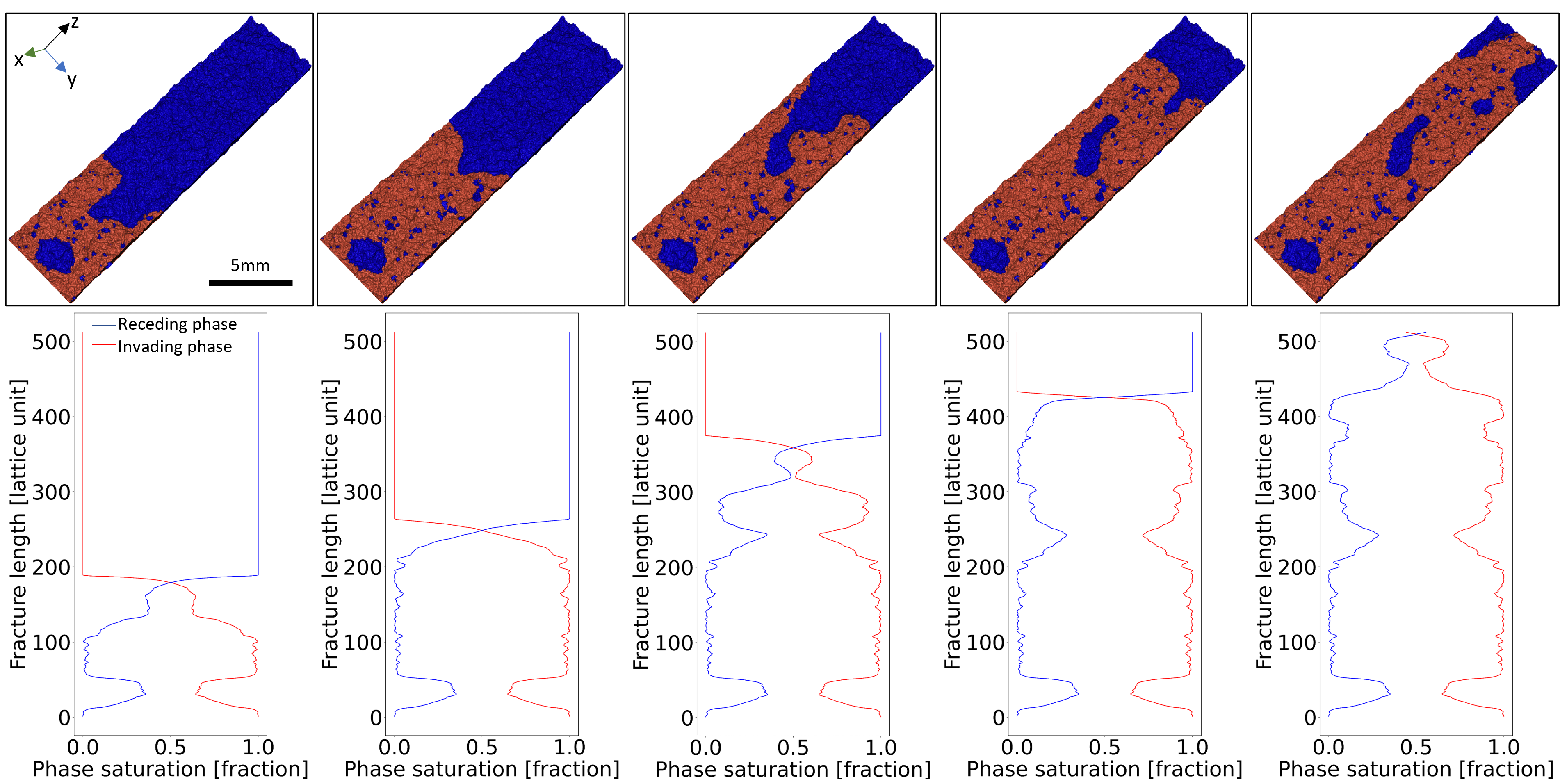}
    \caption{Implementation of \pysimfrac to generate multiphase flow data. Upper row shows snapshots of increasing saturation of the red phase inside the fracture. Lower row shows corresponding saturation profiles of the red and blue phases along the length of the fracture. The scale is adapted from the fracture mean aperture (0.584 mm) in \cite{karpyn2007visualization}}
    \label{Fig_4_1_2}
\end{figure}

The same fracture properties in the MP-LBM integration (Section 3.1) were used to generate a fracture for simulation with MF-LBM (Figure \ref{Fig_4_1_2}). The initial condition in the fracture began with 100\% occupancy by the blue phase, which was also the wetting phase. The contact angle was set to 50\textdegree. Fluid viscosity ratio was set as 1.0, while the capillary number was set at 10E-4. The red phase (non-wetting phase) was introduced from the bottom. With the snapshots shown in Figure \ref{Fig_4_1_2}, we see an increase in the occupancy of the red phase as injection proceeds. We also estimated the corresponding saturation of both phases for each time step. At the last time step, we notice the majority of the fracture is occupied by the red phase with as saturation of 87.6\%.

\subsection{Integration with dfnWorks: Three-Dimensional Discrete Fracture Network}

{\sc dfnWorks} is an open source three-dimensional discrete fracture network (DFN) modeling suite~\citep{hyman2015dfnworks}.
In a 3D DFN model, fractures are represented as a network of intersecting planes. 
The size, shape, orientation, and other hydrological properties are sampled from distributions whose parameters are determined from a site characterization, cf.~\cite{viswanathan2022from} for a comprehensive discussion of DFN modeling approaches.
Once the network is produced, as computational mesh representation is generated, on which flow and transport can be resolved~\citep{hyman2014conforming,krotz2022maximal}.
Details of {\sc dfnWorks} in terms of algorithms and various applications can found in \cite{hyman2015dfnworks}.

\begin{figure}
    \centering
    \includegraphics[width = 0.8\textwidth]{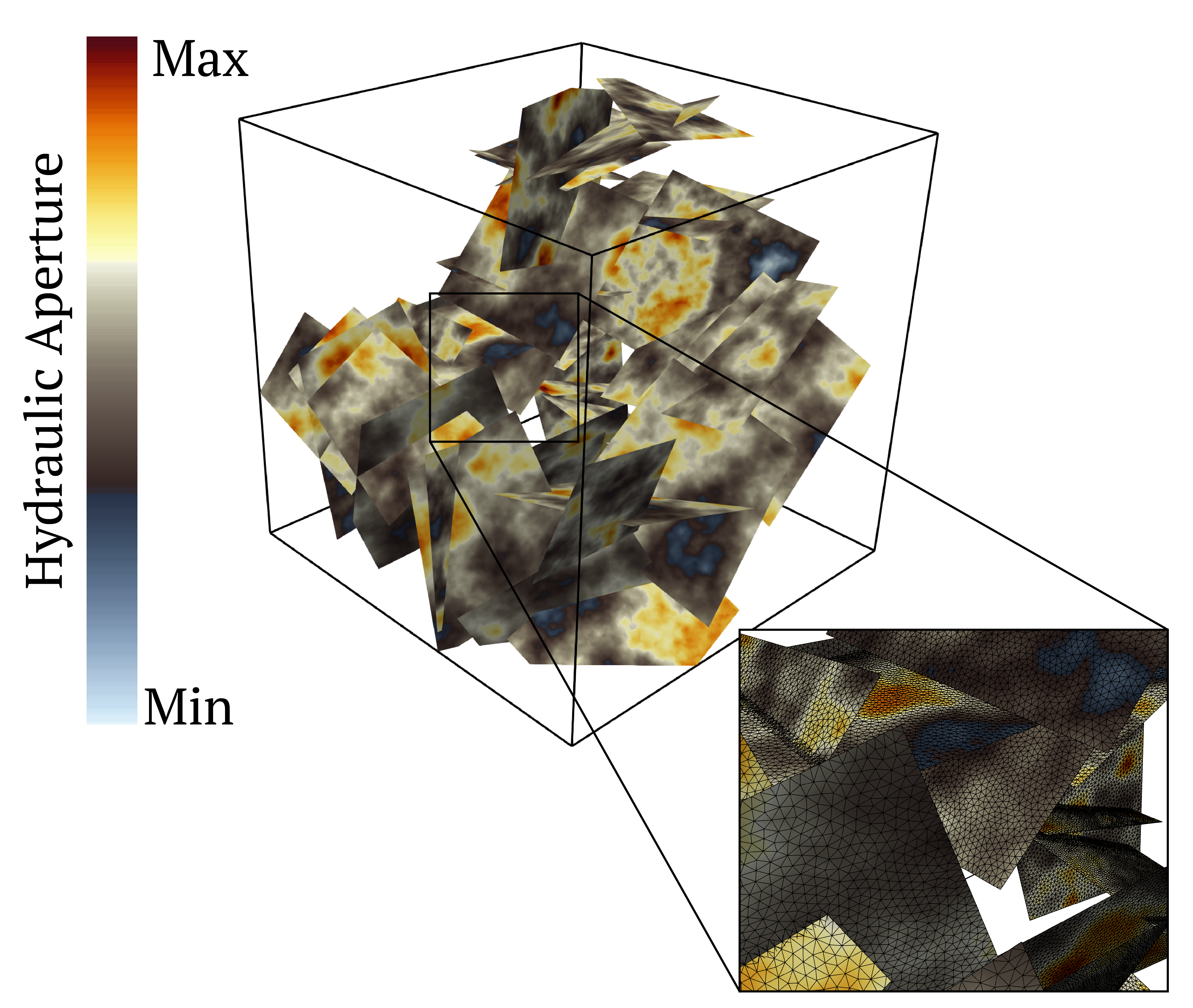}
    \caption{A three-dimensional discrete fracture network generated with {\sc dfnWorks} that includes internal aperture variability generated using the \pysimfrac spectral method} 
    \label{fig:dfn}
\end{figure}

A key capability of {\sc dfnWorks} is the ability to include variable aperture values into a 3D DFN simulation, e.g.,~\cite{karra2015effect,frampton2019advective,makedonska2016evaluating,hyman2021scale}.
We developed a handshake between \pysimfrac and {\sc dfnWorks} to map \pysimfrac generated aperture fields directly onto {\sc dfnWorks} fractures. 
An example of these is shown in Fig.~\ref{fig:dfn}. 
The network is composed of thirty-eight two meter square fractures in a 10 meter cube. 
Each fracture has a unique aperture field generated using the \pysimfrac spectral method. 
Each node in the mesh is assigned an aperture value from a \pysimfrac fracture (see inset).

\section{Conclusions}

The study of complex fractures has important applications in many research areas within the geosciences.  Here, we present a new synthetic fracture generation library and analysis toolkit which allows for the investigation of wide range of fracture properties.  Implemented in python and available open-source, \pysimfrac makes it significantly easier to create and analyze realistic fractures for a wide range of research applications.  In addition the integration with open-source simulation codes such as dfnWorks, MP-LBM, and MF-LBM makes fracture network generation and direct numerical simulation fast and approachable.  The ability to easily create and simulate upon fractures which span the expected variety from nature should yield important findings in a range of disciplines. 

\section{Acknowledgments}

Research presented in this article was supported by the Laboratory Directed Research and Development program of Los Alamos National Laboratory under project number XXKF00 and has been designated with the Los Alamos Unlimited Release number LA-UR-23-26998. J.S. would like to thank the Center for Nonlinear Studies for support.
J.D.H acknowledges support from the Department of Energy (DOE) Basic Energy Sciences program (LANLE3W1) for support. 
This work has been partially funded by the Spent Fuel and Waste Science and Technology (SFWST) Campaign of the U.S. Department of Energy Office of Nuclear Energy.
The views expressed in the article do not necessarily represent the views of the U.S. Department of Energy or the United States Government.

\newpage

The source codes are available for downloading at the link:
https://github.com/lanl/dfnWorks

\bibliographystyle{cas-model2-names}
\bibliography{bibliography} 

\end{document}